\documentclass[prl,amsmath,amssymb,twocolumn,superscriptaddress]{revtex4-2}
\usepackage{graphicx,bm,color,amsthm,amsmath,amsfonts,appendix}

\newcommand{\tbm}[1]{\tilde{\bm{#1}}}

\begin{document}
\title{Dirac electron under periodic magnetic field: \\ Platform for  fractional Chern insulator and generalized Wigner crystal}

\author{Junkai Dong}
\affiliation{Laboratory of Atomic and Solid State Physics, Cornell University, Ithaca, NY 14853}
\affiliation{Department of Physics, Harvard University, Cambridge, MA 02138}

\author{Jie Wang}
\affiliation{Center for Computational Quantum Physics, Flatiron Institute, 162 5th Avenue, NY 10010}

\author{Liang Fu}
\affiliation{Department of Physics, Massachusetts Institute of Technology, Cambridge, MA 02139}

\begin{abstract}
We propose a platform for flat Chern band by subjecting two-dimensional Dirac materials---such as graphene and topological insulator thin films---to a periodic magnetic field, which can be created by the vortex lattice of a type-II superconductor. As a generalization of the $n=0$ Landau level, the flat band of Dirac fermion under a nonuniform magnetic field remains at zero energy, exactly dispersionless and topologically protected, while its local density of states is spatially modulated due to the magnetic field variation. In the presence of short-range repulsion, we find fractional Chern insulators emerge at filling factors $\nu=1/m$, whose ground states are generalized Laughlin wavefunctions. We further argue that  generalized Wigner crystals may emerge at certain commensurate fillings under a highly nonuniform  magnetic field in the form of a flux line lattice.   
\end{abstract}

\maketitle
Flat bands provide an ideal venue for exploring strongly correlated electron phenomena due to the vanishing kinetic energy. Of particular interest are systems with topological flat bands, which hold great promise for realizing fractional topological phases of matter. Unlike the continuum Landau levels, the physics of topological flat bands is affected by the periodicity of Bloch electron wavefunctions, which has no analog in quantum Hall systems. Over the last decade, Dirac electron systems have emerged as an appealing platform for engineering flat bands, which culminated in the discovery of magic-angle twisted bilayer graphene~\cite{Andrei:2020aa,Bistritzer12233}. Topological flat bands in the absence of external magnetic field have also been proposed in other moir\'e materials~\cite{Zhen_LF19,Grisha_TBG,Eslam_hierarchy,JieWang_NodalStructure,Devakul:2021aa,Paul_LF21,Jie_HierarchyIdealBands,Ashvin_FamilyIdealBands,2021arXiv211112107L}.

In parallel to the pursuit of magic-angle graphene, a widely studied approach to create flat bands is to couple Dirac electrons to an effective U(1) gauge field. In graphene, it was found that triaxial strain in nanometer sized bubbles induces pseudomagnetic fields of opposite signs in the two Dirac valleys, which leads to Landau level like energy spectrum in the local region~\cite{doi:10.1126/science.1191700}. However, due to the bounded nature of strain field it is impossible to sustain a constant pseudomagnetic field over a large area. Subsequent studies turned to the case of a periodically modulated strain field~\cite{Tang:2014aa,Esquinazi:2014aa,PhysRevB.93.195126,Peltonen:2020aa,PhysRevB.102.035425}, which are experimentally realized by misfit dislocation arrays in topological crystalline insulator (TCI) interfaces~\cite{PhysRevLett.88.015507,Zeljkovic:2015aa}, or by buckled graphene superlattices~\cite{Mao:2020aa}. A periodic strain (= gauge) field creates a pseudomagnetic field that is spatially alternating and averages to zero. The resulting band structure features flat dispersion near the Dirac point, leading to diverging density of states which may explain the observed high-temperature superconductivity in TCI and graphite~\cite{Tang:2014aa,PhysRevB.98.054515,Esquinazi:2014aa,Wang_LF21}.

In this work, we propose a simple and feasible approach to engineer flat Chern bands in Dirac materials and realize topological and correlated electron phases. The idea is to simply subject Dirac electrons to a spatially varying magnetic field with nonzero mean. This can be experimentally achieved by placing the Dirac material---such as graphene or topological insulators---above a type-II superconductor  (e.g., niobium). When a magnetic field is applied, the vortex lattice of the superconductor generates a spatially modulated magnetic field. Such nonuniform magnetic field acting on two-dimensional Dirac fermions produces an exactly flat Chern band at zero energy, which is a generalization of the well-known $n=0$ Landau level in a uniform field. We explicitly derive the flat band wavefunction and reveal the effect of the magnetic field modulation on the local density of states. We show analytically that in the presence of short-range electron repulsion, fractional Chern insulators (FCIs) emerge at filling factors $\nu=1/m$ ($m$ is an integer), whose ground states are generalized Laughlin wavefunctions. We argue that {\it incompressible} Wigner crystal states may be found at other fractional fillings that are commensurate with the magnetic field lattice. Our flat band system thus promises tuning the quantum phases between FCIs and generalized Wigner crystals.

There are essential differences between Dirac electron's flat bands created by a modulated magnetic field and by a periodic strain field~\cite{Tang:2014aa,PhysRevB.102.035425}. The strain induced pseudomagnetic field couples oppositely to the two Dirac valleys due to time-reversal symmetry, whereas the real magnetic field breaks time-reversal symmetry and couples equally to the two valleys, producing identical Chern bands. Moreover, the spatially alternating pseudomagnetic field of {\it zero} mean does not open up a gap at the Dirac point or generate isolated minibands. In contrast, 2D Dirac fermion coupled to a periodic magnetic field of {\it nonzero} mean value necessarily exhibits a zero-energy flat Chern band that is topologically protected~\cite{Aharonov_Casher79} and generally separated from higher bands by a finite energy gap.

\begin{figure}
    \centering
    \includegraphics[width=1.0\linewidth]{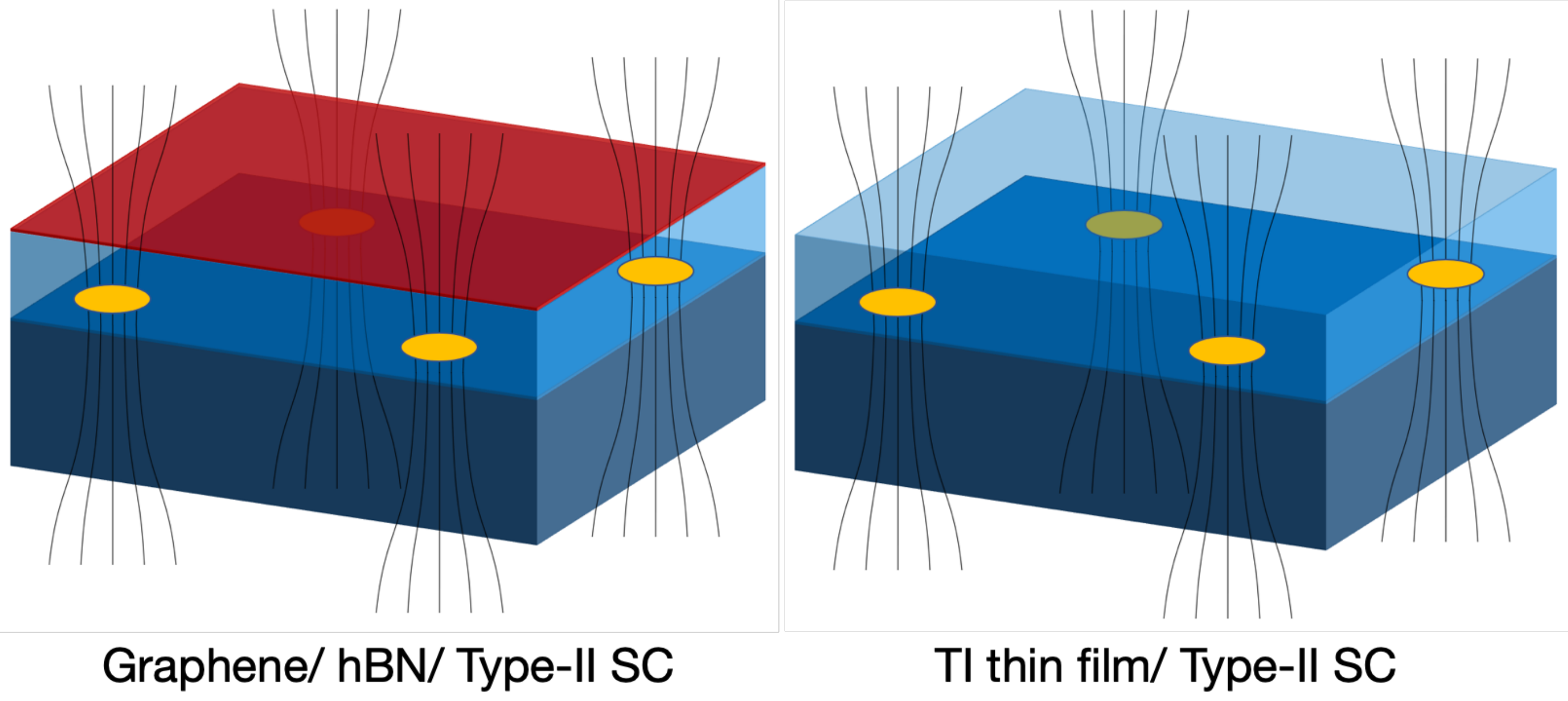}
    \caption{The proposed experimental set up. Left panel: the hetero-structure of a graphene layer (red), hBN (light blue) and type-II superconductor (dark blue), where the hBN prevents the coupling of graphene electron to superconductor and tunes the field strength. Right panel: topological insulator thin film (light blue) and type-II superconductor (dark blue). The vortex lattice is represented by yellow dots.}\label{fig1}
\end{figure}

\emph{Flat Chern band.}---
We consider the energy spectrum of 2D Dirac fermion under a nonuniform magnetic field $B(\bm r)=B+\tilde B(\bm r)$, where $B \neq 0$ is a uniform background part, and $\tilde B$ is the spatially oscillating part of zero mean. In the case of a periodic magnetic field, such as field created by a vortex lattice, the periodicity of $\tilde B$ is set by two primitive lattice vectors  $\bm a_{1,2}$. Denoting the Dirac Fermi velocity as $v_F$, the one-particle Hamiltonian reads:
\begin{eqnarray}
H &=& v_F\sum_{a=x,y}\sigma^a\cdot\left[-i {\hbar}\partial_a - e{A}_a(\bm r) \right],\nonumber\\\label{model}
&=& -\sqrt2 \hbar v_F\left(\begin{matrix} 
& i\partial + \frac{e}{\hbar}A^*(\bm r)\\ 
i\bar\partial + \frac{e}{\hbar}A(\bm r) &  
\end{matrix}\right),
\end{eqnarray}
where $A_a$ is the gauge field associated to the magnetic field $B$. Here $A = (A_x + iA_y)/\sqrt2$ and $\partial = (\partial_x-i\partial_y)/\sqrt2$ are the complex gauge field and holomorphic derivative, respectively.

It is well known that Dirac fermion in a uniform magnetic field exhibits a zero-energy Landau level, with the degeneracy given by the total number of flux quanta $N_\phi$ penetrating the system.
Such zero-mode wavefunction is polarized in the $\sigma_z$ basis and annihilated by the off-diagonal operator $i\bar\partial + eA/\hbar$ in the Hamiltonian: denoting the zero mode wavefunction as $\Psi = \left[\Phi(\bm r), 0\right]^T$ where $\Phi(\bm r)$ is the standard lowest Landau level (LLL) wavefunction satisfying $\left(i\bar\partial + eA/\hbar\right)\Phi=0$. The form of $\Phi(\bm r)$ depends on the choice of gauge field $A$ and the choice of basis. In the symmetric gauge ${\bm A}(\bm r)=(-By, Bx)/2$ or $A=iB z/2$, the zero-mode wavefunction with angular momentum $n$ is $\Phi(\bm r)= z^n\exp(-|z|^2/2l_B^2)$, where $l_B=\sqrt{\hbar/|eB|}$ is the magnetic length. Alternatively the LLL wavefunction in magnetic translation basis $\Phi_{\bm k}(\bm r)$ can be expressed in terms of the modified Weierstrass sigma function as $\Phi_{\bm k}(\bm r) = \sigma(z-z_k)e^{z_k^*z}e^{-\frac12|z|^2}e^{-\frac12|z_k|^2}$ with $z_k$ being $-i(k_x+ik_y)/\sqrt2$~\cite{haldanetorus1,Jie_MonteCarlo,scottjiehaldane,haldanemodularinv,haldaneholomorphic,JieWang_NodalStructure}.

It is also known that for Dirac fermion in a nonuniform magnetic field, there still exists $N_\phi$ zero modes ~\cite{Aharonov_Casher79,dubrovin_novikov_1980}. This can be heuristically understood in the limit of a slowly varying  magnetic field. In a region near $\bm r$, the local energy spectrum is essentially given by the Landau level spectrum of Dirac fermion under a field $B(\bm r)$: $E_n (\bm r)= v_F \sqrt{2 eB(\bm r)|n|}$ [provided that $B(\bm r)\neq 0$]. While the $n\neq 0$ Landau level energies vary with $B$ and become $\bm r$ dependent, the $n=0$ Landau level is always at $E=0$ regardless of the magnetic field strength. Therefore, the zero-energy modes persist when the field becomes nonuniform, whereas other Landau levels are broadened into a dispersive band. This semiclassical argument is valid when the magnetic length is small compared to the characteristic length scale of the magnetic field modulation. Remarkably, the extensively degenerate zero modes persist for any magnetic field of nonzero average, as we will see below.

For a generic nonuniform magnetic field, the gauge field ${\bm A}(\bm r)$ can be split into two parts: ${\bm A} (\bm r) = {\bm A} + \tilde{\bm A}(\bm r)$, where ${\bm A}$ corresponds to the constant magnetic field component (e.g., $A=iB z/2$ in symmetric gauge) and $\tilde{\bm A}$ is associated with the nonuniform field $\tilde{B}(\bm r)$ of {\it zero} spatial average, i.e., $\int d \bm r \tilde{B}(\bm r) = 0$. Mathematically $\tilde{\bm A}(\bm r)$ is guaranteed to be a bounded function. As a consequence, there exists a {\it bounded} scalar potential $\tilde\phi(\bm r)$ such that $\tilde{\bm A} ={\bm  \nabla} \tilde{\phi}$, where $\tilde\phi$ is the solution to  the Poisson equation associated with the nonuniform part of the magnetic field 
\begin{eqnarray}
\tilde B(\bm r) = \nabla^2\tilde\phi(\bm r).
\end{eqnarray}

\begin{figure}
    \centering
    \includegraphics[width=1.0\linewidth]{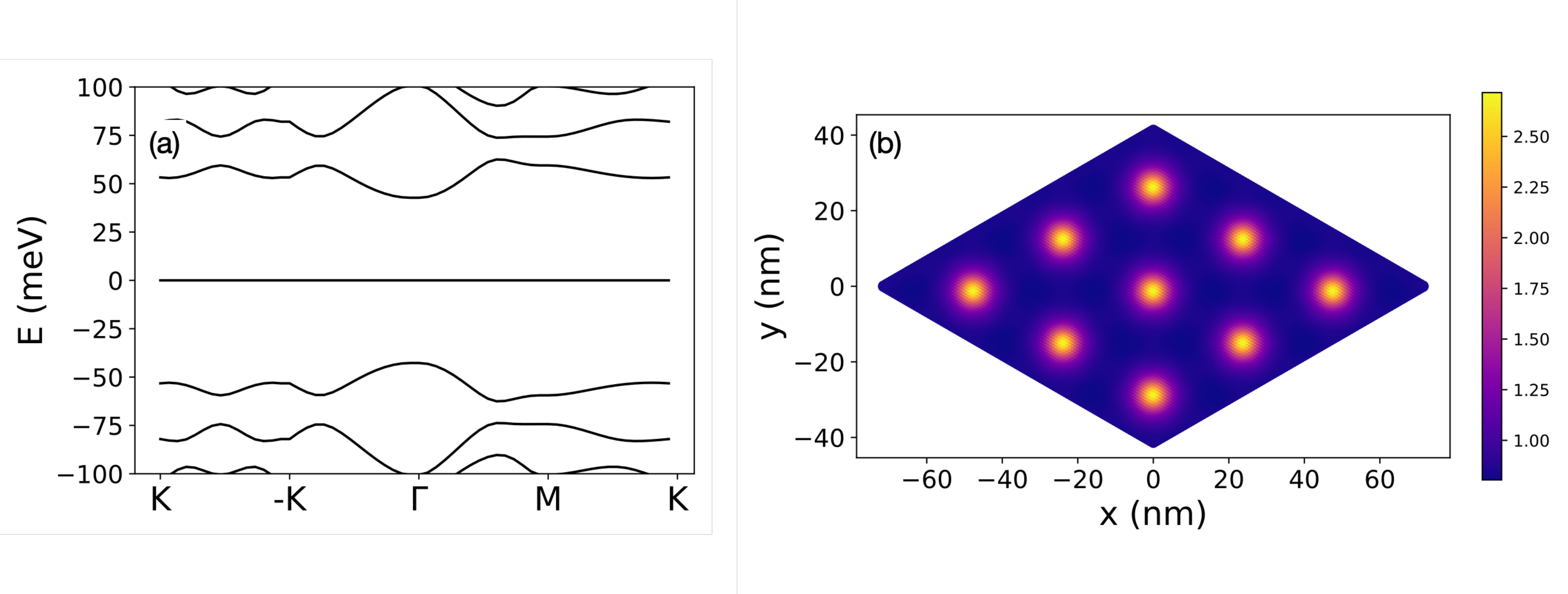}
    \caption{Band structure (left) and LDOS (right) of Dirac fermion under a highly nonuniform magnetic field that varies periodically with the unit cell enclosing one flux quanta ($p=q=1$). This shows the existence of exact flatband, the particle-hole symmetry and the spatially periodic structure of the LDOS.}\label{fig2}
\end{figure}

The zero mode wavefunctions, which should be annihilated by the operator $i\bar\partial+ i B z/2+ i\bar\partial\tilde\phi$ [in unit $e=\hbar=1$], are then simply obtained by multiplying $\exp\left[-\tilde\phi(\bm r)\right]$ to the Landau level wavefunction in the uniform field $B$: 
\begin{equation}
    \Psi(\bm r) = \left[\psi(\bm r),0\right]^T;\quad \psi(\bm r) = \exp\left[-\tilde\phi(\bm r)\right]\Phi(\bm r),\label{generalzeromode}
\end{equation}
where $\Phi(\bm r)$ is the lowest Landau level wavefunction wavefunction.
Note that different choices of gauge and basis lead to different forms of $\tilde\phi(\bm r)$, $\Phi(\bm r)$ and therefore the zero mode wavefunction $\Psi(\bm r)$. However, it does not alter this general structure of the zero mode wavefunction of Dirac fermion under a nonuniform field, without any assumption about its spatial dependence. The effect of magnetic field modulation is fully encoded in the modulation factor $\exp\left[-\tilde\phi(\bm r)\right]$, which leads to a spatial variation of the local density of states of the flat band, as illustrated in Fig.~(\ref{fig2}). We see that by tuning the charge density distribution, magnetic field modulation is a promising tool for tuning quantum phases.

\emph{Generalized Laughlin wavefunctions.---}
While the charge density of the flat band is spatially varying and can be highly nonuniform, we now prove that fractional quantum Hall type states are always the exact ground state as long as short-ranged interactions are considered. This is a consequence of the exact mapping between our flat band zero mode wavefunction $\psi(\bm r)$ and LLL wavefunction $\Phi(\bm r)$, which are simply related by the multiplicative factor $\exp(-\tilde\phi)$ as shown by Eq.(\ref{generalzeromode}). For any given $N$-particle wavefunction in the LLL, multiplying it by such factors for all particles is guaranteed to produce a many-body wavefunction in the flatband Hilbert space.

We now introduce and consider a generalization of the $\nu=1/m$ Laughlin wavefunction to the case of nonuniform magnetic fields:
\begin{equation}
    \Psi_m(\{\bm r\}) = \prod_{i=1}^Ne^{-\tilde\phi(\bm r_i)}\times\prod_{i<j}^{N}\left(z_i-z_j\right)^m\prod_{i=1}^{N}e^{-|z|^2/2l_B^2}.\label{laughlin}
\end{equation}

An important property of model FQH wavefunctions (including the Laughlin) is that they are exactly annihilated by short ranged interactions projected to the LLL~\cite{Haldane_hierarchy,TrugmanKivelson85}. For Laughlin $\nu=1/3$ state, it is the exact ground state of $v_1=\sum_{i<j}\delta''(\bm r_i-\bm r_j)$ interaction. This is seen by examining two particles $z_{1,2}$: when they approach each other the Laughlin wavefunction vanishes as $|z_1-z_2|^3$, a power faster than what is required by Fermi-Dirac statistics. This two-particle short ranged behavior implies the $\nu=1/m$ Laughlin wavefunction   is a zero-energy eigenstate of positive interaction energies $v_k$ for any two particles with relative angular momentum $k<m$. This reasoning can be applied straightforwardly to prove an exact property of our flat band wavefunction. Since the one-body factor  $\exp(-\tilde\phi)$ does not change the $|z_1-z_2|^m$ behavior of the many-body wavefunction as $z_1 \rightarrow z_2$, our generalized Laughlin wavefunction $\Psi_m$ remains exactly zero energy ground state for short ranged repulsions.

\emph{Periodic magnetic field and ideal flat Chern band.---}
Our discussion so far are general without assuming the spatial profile of the magnetic field. From now on we consider magnetic fields with periodic spatial modulation. Such periodic magnetic field can be naturally realized in the vortex lattice of type II superconductors, which we will study in detail later. We start with a general discussion of a periodic magnetic field $B(\bm r) = B(\bm r+\bm a_{1,2})$ where $\bm a_{1,2}$ are primitive lattice vectors. The number of flux quanta per unit cell $\nu_f$ can be rational or irrational. In the rational case with $\nu_f=p/q$ ($p,q$ are co-prime integers), a magnetic unit cell consisting of $q$ lattice unit cells can be found, which encloses $p$ flux quanta.

The magnetic field created by the vortex lattice of a superconductor corresponds to the rational case with $p=1$ and $q=2$, as a single vortex carries half a flux quanta $h/2e$. In this case, the total number of flux quanta threaded through the Dirac material is half the number of vortices or lattice sites: $N_\phi = \frac{1}{2} N_s$. The $N_\phi$ degenerate zero modes can now be labeled by their magnetic translation momentum $\bm k$: $\psi_{\bm k}(\bm r) = \exp\left[-\tilde\phi(\bm r)\right]\Phi_{\bm k}(\bm r)$. Note that the area of the magnetic Brillouin zone is half the area of the lattice Brillouin zone. We observe that the zero mode wavefunction in $k$ space has the universal form of the $C=1$ flat band with ideal quantum geometry~\cite{JieWang_exactlldescription,Grisha_TBG2}. Compared to other settings for such ideal flat band such as chiral twisted bilayer graphene~\cite{Grisha_TBG,Grisha_TBG2,Jie_HierarchyIdealBands,Ashvin_FamilyIdealBands,FCI_TBG_exp,DanParker21} and Kapit-Mueller models~\cite{Kapit_Mueller,junkaidong20,ModelFCI_Zhao}, our platform of Dirac material under periodic magnetic field is experimentally feasible and relatively straightforward to implement.

\emph{Experimental realization.---}
We now discuss the experimental feasibility of realizing the flat Chern band of Dirac electron using the magnetic field that penetrates through a type-II superconductor. Under an external field $B_{c1}<B<B_{c2}$, vortices in the superconductor usually form a triangular lattice, whose lattice constant $a$ is directly related to the magnetic length $a=\sqrt{h/(\sqrt{3}eB)}=\sqrt{2\pi/\sqrt3}l_B\approx1.9l_B$. In Cartesian coordinate, the primitive vortex lattices are $\bm a_1=a\left(\sqrt3/2,1/2\right)$, $\bm a_2=a\left(-\sqrt3/2,1/2\right)$ and we denote the corresponding reciprocal lattice vector as $\bm g_{1,2}$. Following Ref.~(\onlinecite{Goren:1971aa}), the file profile is determined by the Maxwell equations as well as the boundary conditions at the interface, i.e.,  the $z=0$ plane: 
\begin{eqnarray}
z>0: ~&&\quad\nabla^2\bm B = 0,\label{eqfield1}\\
z<0: ~&&\quad\bm B - \lambda^2\nabla^2\bm B = \sum_i\phi_0\delta(\bm r-\bm r_i)\hat{z},\label{eqfield2}
\end{eqnarray}
where $\lambda$ is the penetration depth of the superconductor, $\phi_0=h/2e$ is the quantized magnetic flux carried by  a single vortex. Solving Eqn.~(\ref{eqfield1}) and Eqn.~(\ref{eqfield2}) with matching boundary condition yields the solution of the field outside the bulk superconductor:
\begin{eqnarray}
    B(\bm r) &=& B + \sum_{\bm g\neq 0}B_{\bm g}\exp\left(-|\bm g|z+i\bm g\cdot\bm r\right),\label{fileprofile1}\\
    B_{\bm g} &\approx& B/\left(1+\lambda^2|\bm g|^2\right).\label{fieldprofile2}
\end{eqnarray}

\begin{figure}
    \centering
    \includegraphics[width=1\linewidth]{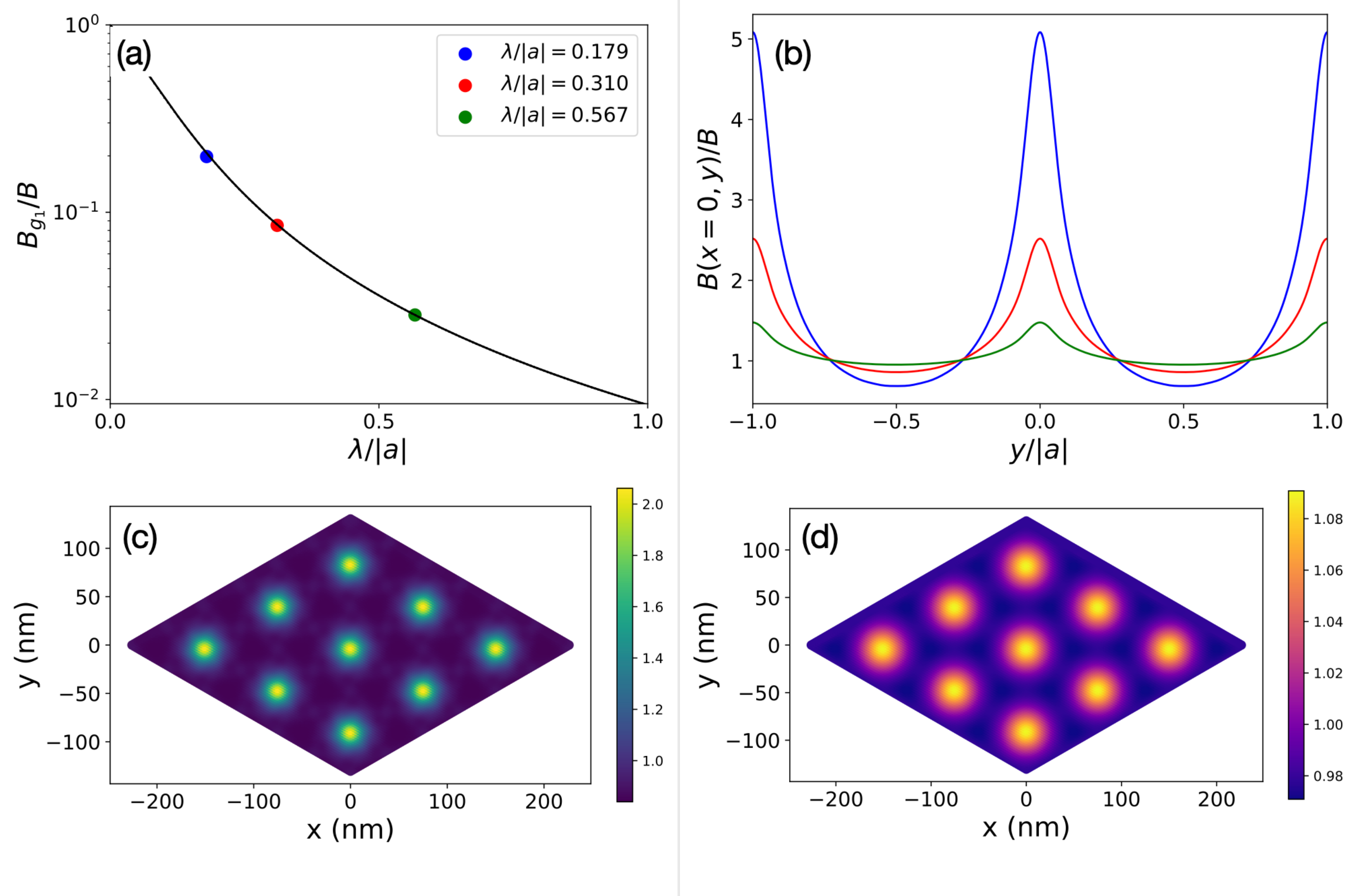}
    \caption{(a) The dependence of the first oscillation ratio $B_{\bm g_1}/B_0$ on the dimensionless ratio $\lambda/|a|$. (b) The spatial field profile at the superconductor surface along $x=0$ plotted as a function of $y$. The $B(x,y)/B$ is only determined by the ratio $\lambda/|a|$. In the plots the penetration depth $\lambda$ is fixed at $27$ nm; the field strengths for the blue, red and green lines are 0.1, 0.3 and 1.0 T respectively whose corresponding $\lambda/|a|$ are found in the legend of (a). The (c) and (d) are the 2D plot of $B(x,y)/B$ and the LDOS $\exp[-2\tilde\phi(x,y)]$ for $B$ at $0.3$T.}\label{fig3}
\end{figure}

The solution shows that the oscillatory part of the magnetic field decay exponentially with the distance $z$ from the superconductor surface, with a decay length determined by the in-plane wave-vector. Therefore the largest contribution comes from the first reciprocal lattice vectors $B_1\equiv B_{\bm g_1}$. The ratio $B_1/B$ is determined by the ratio of the penetration length and vortex spacing $\lambda/a$, as shown in Fig.~\ref{fig3}. At small field $B\ll B_{c2}$ where $a\gg \lambda$, the magnetic field is highly non-uniform as it is confined to flux tubes of size $\lambda$ that are well separated. As the magnetic field increases, these flux tubes start to overlap, which reduces the magnetic field modulation.  Therefore, in order to achieve a significant magnetic field modulation, it is preferable to use type-II superconductors with a small penetration length.

As an example, the superconductor niobium (Nb) has $\lambda=27$nm. For Nb under an external field $B=0.3$T  which corresponds to $\lambda/a=0.3$, the magnetic field distribution near the surface is highly nonuniform: the maximum field at the vortex center is more than twice the minimum field. As a result, the flat band LDOS is periodically modulated by a considerable amount: the modulation factor $\exp\left[-2\tilde \phi(x,y)\right]$ varies by about $10\%$ as shown in Fig.~\ref{fig3}. The challenge with niobium is that its critical field $B_{c2}$ is only around $1$T, which places constraints on the energy scale for quantum Hall states. Other superconductors such as NbSe$_2$ and MgB$_2$ have large $B_{c2}$, but they have a large penetration length more than $100$ nm, which greatly weakens the magnetic field modulation. On a positive note, high-mobility Dirac materials including graphene and topological insulators (Bi$_2$Se$_3$, Sb$_2$Te$_3$ and HgTe) exhibit quantum Hall effect down to very low fields, as small as $0.1$T for graphene and HgTe. Moreover, fractional quantum Hall states have been observed in graphene under a uniform magnetic field as small as $1$T. It will be interesting to search for fractional Chern insulators in graphene under a periodic magnetic field.

\emph{Dirac fermion in magnetic flux lattice.---}
New physics arises when the vortex spacing is much larger than the penetration length, i.e., $a\gg \lambda$. In this case, the magnetic field is confined into a triangular lattice of flux tubes, where the field at vortex center is much larger than the average field $B$. As a result, the zero-energy wavefunction becomes extremely localized at the vortex center. A natural basis for the flat band in such a flux lattice is given by: 
\begin{eqnarray}
\psi_{\bm R}({\bm r})= e^{-\tilde{\phi}({\bm r})} \exp(-|z - z_{\bm R}|^2/2l_B^2),\label{wignerwf}
\end{eqnarray}
where $\bm R$ and $z_{\bm R}$ denote the Cartesian and complex coordinate of vortex centers (= lattice sites) respectively. The zero-mode wavefunction $\psi_{\bm R}({\bm r})$  mainly consists of a large central peak at $\bm R$ and small satellite peaks at adjacent sites. As $\psi_{\bm R}$ occupies several sites, different wavefunctions centered at nearby sites are spatially overlapping and non-orthogonal. Since the dimension of the flat band Hilbert space is only half the number of lattice sites $N_\phi=\frac{1}{2}N_s$, the set of $\psi_{\bm R}$ over all lattice sites $\bm R$ forms an over-complete basis.

\begin{figure}
    \centering
    \includegraphics[width=0.7\linewidth]{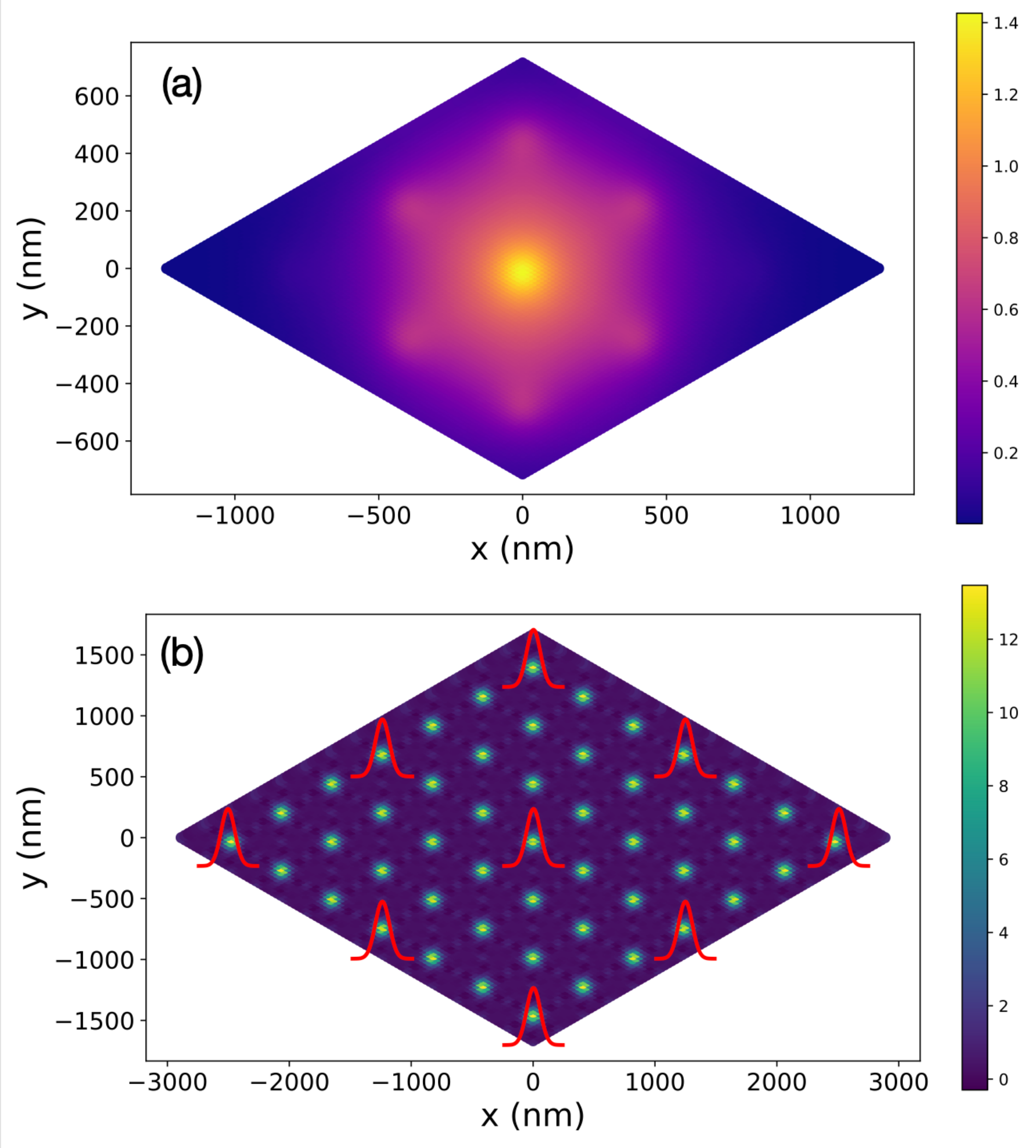}
    \caption{(a) The amplitude of the wavefunction $|\psi_{\bm R=\bm 0}(\bm r)|$ where the satellite peaks are clearly visible. (b) The illustration for the Wigner crystal at filling $\nu=2/9$, where color-map shows the spatial distribution of the magnetic field $B(x,y)/B$ and the red wave-packet illustrates the localized electrons. For both (a) and (b), B is 0.01 T and $\lambda$ = 27 nm.}\label{wignercrystal}
\end{figure}

At low filling fractions where the average distance between electrons is large compared to the lattice constant $a$, Wigner crystal phases are energetically favored to minimize the Coulomb repulsion between neighboring sites. For example, as illustrated in Fig.~\ref{wignercrystal}, a $3\times3$ Wigner crystal is expected at $\nu=2/9$ which corresponds to $1/9$ filling of the flux lattice. A variational wavefunction for this  Wigner crystal is a Slater determinant of localized zero-mode wavefunctions $\psi_{\bm R}$ that form a $3\times3$ superlattice. Unlike the uniform field case, this Wigner crystal state is commensurate with the underlying flux lattice. Due to the locking with the lattice, it is an incompressible state. The energetics of FCIs and Wigner crystal states in the presence of Coulomb interaction will be studied in a future work.

\begin{acknowledgements}
\emph{Acknowledgements.---}
We thank Jinfeng Jia for his interest and valuable feedback. 
This work is supported by a Simons Investigator Award from the Simons Foundation. The Flatiron Institute is a division of the Simons Foundation.
\end{acknowledgements}

\bibliography{ref.bib}

\appendix
\section*{--- Appendix ---}
\section*{Details in solving the Hamiltonian}
The Hamiltonian describing Dirac fermion coupled to gauge field is:
\begin{eqnarray}
    H &=& \hbar v_F\bm\sigma\cdot\left(-i\bm\nabla - \frac{e}{\hbar}\bm A\right),\nonumber\\
    &=& \sqrt{2}\hbar v_F\left(\begin{matrix}&-i\partial+i\frac{eB}{2\hbar}\bar z-\frac{e}{\hbar}\tilde A^*\\-i\bar\partial-i\frac{eB}{2\hbar}z-\frac{e}{\hbar}\tilde A\end{matrix}\right),\nonumber\\
    &=& \frac{\sqrt2\hbar v_F}{l_B}\left(\begin{matrix}&i\bar a^\dag - \frac{\tilde A^*/l_B}{B}\\-i\bar a - \frac{\tilde A/l_B}{B}\end{matrix}\right),\label{DiracH}
\end{eqnarray}
where we have defined $z=(x+iy)/\sqrt2$, $\partial=(\partial_x-i\partial_y)/\sqrt2$ and $\tilde A = (\tilde A_x + i\tilde A_y)/\sqrt2$. We have taken symmetric gauge for the uniform magnetic field $A_a = -B\epsilon_{ab}r^b/2$. The ladder operators are,
\begin{equation}
    \bar a = l_B\bar\partial + z/(2l_B),\quad\bar a^\dag = -l_B\partial + \bar z/(2l_B),
\end{equation}
where we have chosen convention $e,B<0$ thus $eB>0$, so the magnetic length in this convention is $l_B^{-2} = eB/\hbar$. Suppose $\tilde\phi$ as the scalar potential,
\begin{equation}
    \tilde A = i\bar\partial\tilde\phi,\quad\tilde B = 2\partial\bar\partial\tilde\phi,
\end{equation}
the zero mode equation for the flat band $\Psi = \left[\psi, 0\right]^T$ is,
\begin{equation}
    0 = \left[-i\bar a - \frac{\tilde A}{l_BB}\right]\psi = l_B\left[-i\bar\partial + \frac{z}{2l_B^2} - \frac{e}{\hbar}\tilde A\right]\psi,
\end{equation}
which implies if $\Phi$ is the zero mode solution under a uniform magnetic field then the zero mode wavefunction with nonuniform field will be:
\begin{equation}
    \psi = \Phi\times\exp\left(-e\tilde\phi/\hbar\right).
\end{equation}

To solve the spectrum we first of all neglect $\tilde A$ and discuss the uniform magnetic field. The Hamiltonian without $\tilde A$ is:
\begin{eqnarray}
    H_0 &=& \frac{\sqrt2\hbar v_F}{l_B}\left(\begin{matrix}&i\bar a^{\dag}\\-i\bar a\end{matrix}\right),\label{append_def_H0}\\
    &=& \frac{\sqrt2\hbar v_F}{l_B}\left(i\sqrt{n+1}\right)c^{\dag}_{0,n+1,\bm k}c_{1,n,\bm k},\nonumber\\
    &-& \frac{\sqrt2\hbar v_F}{l_B}\left(i\sqrt{n}\right)c^{\dag}_{1,n-1,\bm k}c_{0,n,\bm k},\nonumber
\end{eqnarray}
where $c^{\dag}_{s,n,\bm k}$ creates an electron of sublattice $s=0,1$, Landau level index $n=0,...,$ and momentum $\bm k$. The momentum $\bm k$ labels the boundary condition under integer flux insertion. More formally, we define the magnetic translation group $t(\bm q)=\exp\left(i\bm q\cdot\bm R\right)$ where $\bm R$ is the guiding center satisfying $[R^a,R^b]=-i\epsilon^{ab}l_B^2$. The cyclotron motion $\bar{\bm R}\equiv\bm r-\bm R$ corresponds to inter Landau level transitions. The $\bm k$ labels the boundary condition which is the eigenvalue of the periodic magnetic translation $t(\tbm g)|\bm k\rangle = \eta_{\bm b}e^{i\tbm g\times\bm k}|\bm k\rangle$ where $\eta_{\tbm g}=-1$ if $\tbm g/2$ is not a lattice point and $+1$ if otherwise. Here $\tbm g$ is a magnetic reciprocal lattice vector. A generic magnetic translation group element acts as $t(\bm q)|\bm k\rangle = e^{\frac{i}{2}\bm q\times\bm k}|\bm k+\bm q\rangle$. The fluctuation part $\tilde H$ takes the form,
\begin{eqnarray}
    \tilde H &=& \frac{\sqrt2\hbar v_F}{l_B}\left(\begin{matrix}&-\frac{\tilde{A}^*/l_B}{B}\\-\frac{\tilde{A}/l_B}{B}&\end{matrix}\right),\nonumber\\
    &=& \frac{\sqrt2\hbar v_F}{l_B}\left(\begin{matrix}&i\frac{el_B}{\hbar}\partial\tilde\phi^*\\-i\frac{el_B}{\hbar}\bar\partial\tilde\phi\end{matrix}\right),
\end{eqnarray}
where in getting the second equation we used the relation between the scalar and vector gauge connections $\tilde A_a = -\epsilon_{ab}g^{bc}\partial_c\tilde\phi$ where in our case the metric $g^{ab}=\delta^{ab}$ as seen from Eqn.~(\ref{DiracH}). In complex coordinate representation we have $\tilde A = i\bar\partial\tilde\phi$ mentioned before. We now want to project the $\tilde H$ into the basis $|s,n,\bm k\rangle$. Since the nonuniform field does not mix magnetic translation boundary conditions (because the wave-vector of $\tilde A$ is commensurate), only $\bm k-$diagonal matrix element is relevant. We introduce the Fourier modes of the $\tilde A_a(\bm r)$ as,
\begin{equation}
    \tilde A_a(\bm r) = \sum_{\tbm g}\tilde A_a(\tbm g)\exp(i\tbm g\cdot\bm r)
\end{equation}
where $\tilde g$ is the reciprocal lattice for the magnetic unit cell. We define the following matrix elements:
\begin{eqnarray}
    && \mathcal{M}_{s,s';n+m,n}(\bm k) \equiv \langle s,n+m,\bm k|\tilde H|s',n,\bm k\rangle,\\
    &=& -\frac{\hbar v_F}{l_B}\sum_{a=x,y}\sum_{\tbm g}\sigma^a_{ss'}~\frac{\tilde A_a(\tbm g)}{Bl_B}~f^{\bm k}_{n+m,n}(-\tbm g),\nonumber
\end{eqnarray}
where $f^{\bm k}$ is the diagonal element of the following,
\begin{eqnarray}
    f^{\bm k\bm k'}_{n+m,n}(-\tbm g) &\equiv& \langle n+m,\bm k|e^{i(\bm k-\bm k'+\tbm g)\cdot\bm r}|n,\bm k'\rangle,\\
    &=& \eta_{\tbm g}e^{-\frac{i}{2}(\bm k+\bm k')\times\tbm g}e^{\frac{i}{2}\bm k\times\bm k'}g_{n+m,n}(\bm k-\bm k'+\tbm g),\nonumber
\end{eqnarray}
and $g$ satisfies $g_{n+m,n}(\bm q) = g^*_{n,n+m}(-\bm q)$ and is defined in below:
\begin{eqnarray}
    g_{n+m,n}(\bm q) &\equiv& \langle n+m|e^{i\bm q\cdot\bar{\bm R}}|n\rangle,\label{LLformfactor}\\
    &=& \sqrt\frac{n!}{(n+m)!}L^m_n(qq^*)(iq)^m\exp\left(-\frac12qq^*\right).\nonumber
\end{eqnarray}

\section*{Field Profile near the Surface of Superconductor}
\subsubsection{Vortex lattice and magnetic lattice}
The above discussion is generally true for any geometry of magnetic lattice. In this section, we specify to the case of vortex lattice. Since the magnetic unit cell contains area $2\pi l_B^2$, twice the area of the vortex lattice, without loss of generality we set the lattice vectors for the magnetic lattice as $\tbm a_{1,2}$ which is related to the lattice vectors of the vortex lattice $\bm a_{1,2}$ via:
\begin{equation}
    \tbm a_1 = 2\bm a_1,\quad \tbm a_2 = \bm a_2;\quad \tbm g_1 = \bm g_1/2,\quad \tbm g_2 = \bm g_2,
\end{equation}
where $\bm a_{1,2}$ spans a standard triangular lattice. The $\bm g_{1,2}$ and $\tbm g_{1,2}$ are the reciprocal lattice vectors for the vortex lattice and magnetic lattice, respectively. The area of the vortex lattice unit cell $2\pi S$ is:
\begin{equation}
    S \equiv \frac{\sqrt3}{2}a^2 = \pi l_B^2 = \frac{\phi_0}{|B|},\label{deffluxquanta}
\end{equation}
where $\phi_0 = h/(2e)$ is the superconducting magnetic flux quanta. The concrete expression of lattice vectors are listed below:
\begin{eqnarray}
    \bm a_1 &=& a\left(\frac{\sqrt3}{2}, \frac12\right),\quad\bm a_2 = a\left(-\frac{\sqrt3}{2}, \frac12\right);\nonumber\\
    \bm g_1 &=& \frac{4\pi}{\sqrt3a}\left(\frac12, \frac{\sqrt3}{2}\right),\quad\bm g_2 = \frac{4\pi}{\sqrt3a}\left(-\frac12, \frac{\sqrt3}{2}\right),\nonumber
\end{eqnarray}
and the magnetic lattice vectors are simply $\tbm a_1=2\bm a_1$, $\tbm g_1=\bm g_1/2$ and $\tbm a_2 = \bm a_2$, $\tbm g_2=\bm g_2$. The scalar potential $\tilde\phi$ and the gauge potential are periodic in vortex lattice. By using $\tilde A_{\bm g}/Bl_B = -gl_B\times e\tilde\phi_{\bm g}/\hbar$, the Fourier transform for the scalar potential and the complex gauge potential are,
\begin{eqnarray}
    \tilde\phi(\bm r) &=& \sum_{m,n}\tilde\phi_{m\bm g_1+n\bm g_2}\exp\left[i(m\bm g_1+n\bm g_2)\cdot\bm r\right],\nonumber\\
    &=& \sum_{m,n}\tilde\phi_{m\bm g_1+n\bm g_2}\exp\left[i(2m\tbm g_1+n\tbm g_2)\cdot\bm r\right];\nonumber\\
    \frac{\tilde A(\bm r)}{Bl_B} &=& \sum_{m,n}\frac{\tilde A_{m\bm g_1+n\bm g_2}}{Bl_B}e^{i(2m\tbm g_1+n\tbm g_2)\cdot\bm r},\nonumber\\
    &=& -\frac{el_B}{\hbar}\sum_{m,n}(mg_1+ng_2)\tilde\phi_{m\bm g_1+n\bm g_2}e^{i(2m\tbm g_1+n\tbm g_2)\cdot\bm r}.\nonumber
\end{eqnarray}

\subsubsection{Derivation of the magnetic field profile near the superconductor surface}
We review the derivation of the magnetic field profile near the surface of the superconductor following Ref.~(\onlinecite{Goren:1971aa}). The field profile is constrained by the Laplacian equation. They are solved from the following differential equations for the magnetic field inside and outside the superconductor:
\begin{eqnarray}
\text{(outside)}&&\quad\nabla^2\bm B = 0,\label{solution_out}\\
\text{(inside)}&&\quad\bm B - \lambda^2\nabla^2\bm B = \sum_i n\phi_0\delta(\bm r-\bm r_i)\hat{z}.\label{solution_in}
\end{eqnarray}

Solving the above equation with the standard interface boundary condition yields the magnetic field profile. For a 3D superconductor of width $d$ and penetration depth $\lambda$, the field at vertical distance $z$ is,
\begin{eqnarray}
    \tilde B(\bm r, z) &=& B + \sum_{\bm g}B_{\bm g}(z)\cos(\bm b\cdot\bm r),\\
    B_{\bm g}(z) &=& B_{\bm g}\exp\left[-|\bm g|(|z|-d/2)\right],\\
    B_{\bm g} &=& \frac{B/(\lambda K_{\bm g})^2}{1+\frac{|\bm g|}{K_{\bm g}}\coth\left(\frac{K_{\bm g}d}{2}\right)},
\end{eqnarray}
where $d/2$ is location of the surface, $K_{\bm g}^2 = \bm g^2 + 1/\lambda^2$. The gauge connection is $\tilde A_{\bm g,a} = i\epsilon_{ab}g^{bc}g_cB_{\bm g}/|\bm g|^2$. Written in terms of dimensionless quantities, the gauge connection and the scalar potential are:
\begin{eqnarray}
    \frac{\tilde A_{\bm g,a}}{Bl_B} &=& i\epsilon_{ab}g^{bc}\frac{g_c/l_B}{|\bm g|^2}\frac{1/(\lambda K_{\bm g})^2}{1+\frac{|\bm g|}{K_{\bm g}}\coth\left(\frac{K_{\bm g}d}{2}\right)},\\
    \frac{e}{\hbar}\tilde\phi_{\bm g} &=& -\frac{B_{\bm g}}{B}\frac{1}{l_B^2|\bm g|^2} = -\frac{1}{l_B^2|\bm g|^2}\frac{1/(\lambda K_{\bm g})^2}{1+\frac{|\bm g|}{K_{\bm g}}\coth\left(\frac{K_{\bm g}d}{2}\right)}.\nonumber
\end{eqnarray}

In the main test we have taken $d\rightarrow\infty$ valid for 3D superconductors.

\end{document}